# Understanding AI Anxiety in the Workplace: A Multimethod Investigation Using Fear Acquisition Theory and the Technology Acceptance Model

Jaroslaw Grobelny[1]*, Mateusz Klakus[1], Kacper Szymański[1], Teresa Chirkowska-Smolak[1]

[1]Faculty of Psychology and Cognitive Science, Adam Mickiewicz University, Poznań, Poland

*Correspondence: Jaroslaw Grobelny, Faculty of Psychology and Cognitive Science, Adam Mickiewicz University, Szamarzewskiego 89AB, 61-478 Poznań, Poland. E-mail: jaroslaw.grobelny@amu.edu.pl. ORCID: 0000-0003-4296-402X

## Abstract

As artificial intelligence (AI) rapidly diffuses and concerns about job displacement intensify, the psychological mechanisms underlying AI job replacement anxiety remain insufficiently understood. Drawing on Integrated Fear Acquisition Theory and the Technology Acceptance Model, the present research investigates whether AI job replacement anxiety can be elicited through vicarious exposure to narratives emphasizing AI-over-human control, and whether perceived usefulness and perceived ease of use of AI moderate this response. Across two studies, we examine AI job replacement anxiety as a response that emerges through vicarious exposure to narratives emphasizing AI agency and human control loss, rather than through direct personal experience of job displacement. Study 1 employed a randomized experiment (N = 316), demonstrating that such exposure increased AI job replacement anxiety. This effect was moderated by perceived usefulness of AI, but not by perceived ease of use, and remained robust after controlling for core self-evaluations. Study 2 (N = 995) replicated the association between perceived AI-over-human control and job replacement anxiety in an observational design and provided convergent evidence for the moderating role of perceived usefulness, supporting the external validity of the findings. Together, the results provide the first causal evidence that perceptual and vicarious processes can trigger AI job replacement anxiety. By shifting attention from structural labor-market conditions to how AI agency is perceived and communicated, this work offers a mechanism-based account of when and why AI-related job fears arise.

## Introduction

The rapid advancement of artificial intelligence (AI) is expected to reshape labor markets substantially, yet experts remain divided on whether AI will create more jobs than it replaces [1,2]. From the job-replacement perspective, AI is projected to displace a large share of the global workforce, with some evidence suggesting this process has already begun [3,4]. Notably, the pace at which technology displaces workers seems to exceed the ability of educational institutions and training programs to adapt adequately [2]. Consequently, many employees report anxiety related to AI-driven job loss, affecting both personal and collective employment prospects [5], including workers from technologically proficient generations [6]. AI's representation in digital media further amplifies this anxiety. Research shows that social media (SM)—particularly short, algorithmically recommended video content—plays a key role in shaping attitudes toward AI [7]. Sentiment and topic analyses consistently reveal predominantly negative AI-related content, emphasizing societal and ethical threats, dehumanization, and concerns that AI could become too powerful to control [7–9]. Crucially, fear of AI's potential consequences has already been linked to adverse outcomes for employee behavior and well-being [10,11]. Together, these trends highlight an urgent need to understand how exposure to AI narratives—particularly those portraying AI as an autonomous agent exerting control over people and replacing human labor—contributes to negative attitudes toward AI and, most notably, to AI-related job replacement anxiety. Understanding how these narratives shape perceptions of AI agency and control is essential for explaining when exposure to the technology itself translates into psychological responses.

AI anxiety has emerged as one of the central psychological attitudes studied in relation to AI, reflecting individuals' apprehension toward the rapid development and societal integration of AI technologies [12,13]. Conceptually, AI anxiety is understood as a multidimensional construct encompassing concerns related to learning demands, job replacement, loss of human control, ethical violations, and broader sociotechnical risks. Empirical research indicates that AI anxiety is closely related to, yet distinct from, earlier constructs such as automation anxiety or STARA awareness, which focus more narrowly on technology-driven job insecurity and physical task automation [10,14,15]. Among the multiple dimensions of AI anxiety, job-replacement anxiety is a core component, capturing forward-looking fears that AI technologies will render human labor obsolete in the workplace [12,13]. This dimension reflects concerns about job loss, increased reliance on AI systems, and the erosion of human skills and autonomy, including the belief that reliance on AI may undermine reasoning abilities or replace human workers altogether. As such, job replacement anxiety constitutes a central manifestation of AI anxiety, grounded in perceptions of AI as a powerful and increasingly autonomous force capable of displacing human roles across occupational domains [16].

Research on the antecedents of AI anxiety is still limited, with theoretical accounts preceding extensive empirical testing. Early work suggests that AI anxiety often reflects inaccurate perceptions of current technologies, including exaggerated future scenarios (e.g., superintelligence) and confusion between computational autonomy and human agency [13,16]. More recent accounts describe AI anxiety as a societal response to the rapid development and integration of AI, encompassing concerns about job security, loss of human control, and social disruption [14]. Available evidence indicates that AI anxiety is associated with labor-market and job-related factors, such as perceived employment threat, skill obsolescence, and work identity, as well as demographic and individual characteristics, including age, education, employment status, technological literacy, and personality [17,18]. Reviews also highlight perceived risks related to AI technologies, including bias, misinformation, privacy concerns, uncontrolled growth, and the erosion of human skills and workplace social interaction [14].

In contrast, considerably less is known about the specific formation of job replacement anxiety. Initial evidence from related constructs, such as fear of automation, links job replacement concerns to exposure to automation technologies, employment in automatable occupations, perceived lack of control, and demographic factors [15]. Importantly, theoretical and empirical work also highlights the role of vicarious exposure, in which job replacement anxiety develops through information about others losing their jobs to AI rather than through direct personal experience [12]. However, important gaps remain in understanding replacement anxiety emergence. Existing evidence is based largely on observational studies [15,e.g., 19–21] and focuses on objective technological characteristics, workforce-related risks, or individual traits. By contrast, relatively little attention has been paid to how AI is perceived, framed, and communicated in media narratives. This represents a critical gap, given the widespread exposure to such content and its capacity to shape perceptions of AI agency and control. Moreover, existing studies have made limited use of broader theoretical models. Furthermore, research tends to emphasize general AI anxiety rather than examine its specific dimensions, which differ substantially in content and psychological meaning.

The present study addresses these limitations by providing the first experimental examination of the emergence of AI job-replacement anxiety, advancing a causal understanding of this phenomenon. Drawing on Integrated Fear Theory [12] and the Technology Acceptance Model [TAM; 22], an experimental study (N = 316) tested whether vicarious exposure to content emphasizing AI agentic control—as opposed to narratives highlighting human control—increases AI job replacement anxiety, and whether this effect is moderated by perceived usefulness and perceived ease of use of AI, while controlling for key individual characteristics of core self-evaluations. To examine the external validity of the experimentally identified effects, a large-scale observational study (N = 995) was subsequently conducted. The present study advances research on AI anxiety by providing the first experimental evidence that vicarious exposure to narratives emphasizing AI agentic control causally increases AI job replacement anxiety. By identifying exposure to information about AI-driven human replacement as a key psychological mechanism, the study moves beyond correlational accounts and clarifies how this specific dimension of AI anxiety emerges. The findings further show that perceived usefulness (but not ease of use) of AI strengthens the effect of vicarious exposure on job replacement anxiety, suggesting that positive evaluations of AI may intensify, rather than alleviate, fears of labor displacement. Finally, by accounting for core self-evaluations and state anxiety, the study demonstrates that AI job replacement anxiety is shaped by situational information exposure rather than stable individual dispositions alone, thereby introducing the first empirically grounded model of AI job replacement anxiety emergence.

Although the precise triggering mechanisms of AI job replacement anxiety remain insufficiently understood, important foundations for their investigation were laid by Li and Huang [12], who proposed distinct emergence pathways for AI anxiety dimensions based on Integrated

Fear Acquisition Theory. Analyzing eight dimensions of AI anxiety, they argued—and subsequently provided empirical support—that direct personal experience of AI-driven job replacement is relatively rare. This assumption remains valid, as current reports and studies typically focus on estimating the proportion of jobs at risk of future replacement or projecting how many jobs may be displaced by AI within specified time horizons [e.g., 23–25]. As a result, job replacement anxiety is most likely to develop through vicarious exposure, defined as observing or learning about others' adverse or threatening experiences with AI rather than experiencing them firsthand.

Importantly, from the perspective of Integrated Fear Acquisition Theory [12], exposure to others losing their jobs to AI can be understood as a specific instance of a broader fear-eliciting mechanism: the perception that AI is gaining control over domains traditionally governed by humans. In this view, job loss is not the primary fear stimulus in itself, but a salient signal of a deeper shift in human–AI relations, namely the transfer of agency from human actors to autonomous systems. Accordingly, the core driver of AI job replacement anxiety may lie in exposure to narratives portraying AI as operating and making decisions independently of human oversight. Such narratives frame AI not merely as a tool, but as an autonomous agent capable of assuming control over tasks, roles, and decisions traditionally reserved for humans, thereby amplifying perceptions of human replaceability.

However, such narratives often conflate decision autonomy with computational autonomy and rely on inaccurate conceptions of technological development, including neglecting incremental and contextual constraints on AI capabilities [16,26]. As a consequence, perceived shifts in agency are unlikely to be experienced directly and are instead acquired through observation of others' discussions, examples, and stories, fitting squarely within the vicarious exposure pathway specified by Integrated Fear Acquisition Theory. This interpretation is consistent with earlier observations by Kim et al. [14], who identified perceived lack of human oversight over uncontrolled AI growth as a potential catalyst of AI anxiety, as well as with initial correlational evidence linking perceptions of AI autonomy to heightened replacement anxiety responses [21]. Finally, this rationale aligns with broader accounts suggesting that AI job replacement anxiety is fundamentally forward-looking, deriving from speculative beliefs about future technological development [12], which frequently emphasize AI gaining control and agency over humanity [27–29]. As such, we hypothesized that:

*H1: Vicarious exposure to content emphasizing perceived AI control, as opposed to human control, increases AI job replacement anxiety.*

Available evidence indicates that not all forms of AI exposure increase anxiety, as demographic, personality, and job-related factors shape individual responses [15,20,30]. Moreover, theories of fear learning suggest that exposure elicits anxiety only when it contains salient threat cues [31–33]. Accordingly, exposure to an AI agency may have a limited impact when it does not convey observable personal consequences or signals that a shift in control is plausible or imminent. Conversely, perceptions of loss of control alone are unlikely to provoke anxiety unless embedded in a context that implies potential human replacement, thereby rendering the threat personally relevant [34,35]. Together, these findings suggest that AI-related anxiety depends not only on exposure per se, but also on factors related to how technological agency is perceived and framed.

A suitable framework for identifying such factors is TAM, a foundational theory in technology adoption research that explains how individuals come to accept and use new technologies [22]. Extensive empirical evidence across diverse domains shows that perceived usefulness and perceived ease of use are robust predictors of attitudes toward technology and related behavioral intentions [e.g., 36–39]. Perceived usefulness refers to the extent to which an individual believes that using a technology will enhance performance, whereas perceived ease of use reflects the belief that using technology will require minimal effort. These constructs not only exert direct effects on attitudes but also interact, as higher perceived ease of use typically increases perceived usefulness, thereby indirectly shaping evaluative responses to technology [19,40,41]. Owing to its explanatory value, TAM has been widely applied in research on AI usage and acceptance [42–44]. Relatedly, prior work has linked technology-related anxiety to TAM constructs; for example, Saadé and Kira [45] modeled computer anxiety as an antecedent of both perceived ease of use and perceived usefulness in the context of online learning systems.

An important distinction, however, is that TAM was originally developed to explain technology use, whereas AI job replacement anxiety concerns individuals' experience of being subject to technological systems. We argue that this distinction becomes blurred in the case of AI. Contemporary AI technologies increasingly rely on profiling, pattern recognition, and automated decision-making to influence behavior without users' explicit awareness, and are often deployed in contexts such as surveillance or social credit systems [46–48]. As a result, individuals may simultaneously be users of AI-based systems and objects of their operation. Under these conditions, TAM constructs may help explain how vicarious exposure to AI-related narratives translates into heightened job replacement anxiety by shaping perceptions of AI agency and its potential consequences. However, in the present framework, perceived usefulness and perceived ease of use are not treated as direct predictors of AI job replacement anxiety, because anxiety is assumed to emerge only when perceptions of AI agency and control are activated. Instead, both perceptions are expected to moderate the effect of vicarious exposure by shaping the extent to which narratives of AI dominance are interpreted as plausible, consequential, and personally relevant. When AI is perceived as highly useful, narratives emphasizing AI dominance are more credible and more likely to be implemented, thereby increasing perceived job-displacement risk. Likewise, high perceived ease of use lowers perceived barriers to deployment, making AI-driven replacement seem more feasible and imminent. Under both conditions, vicarious exposure amplifies anxiety by signaling that large-scale human replacement is plausible.

The present framework therefore proposes that perceived usefulness and ease of use moderate fear acquisition specifically because they shape the credibility and personal relevance of threat cues — the two conditions Integrated Fear Acquisition Theory identifies as necessary for vicarious fear learning. Perceived usefulness signals that AI-driven replacement is consequential and therefore worth fearing; perceived ease of use signals that it is feasible and therefore imminent. In this reading, TAM constructs do not predict anxiety directly but function as appraisal filters determining whether a narrative of AI dominance is encoded as a personally relevant threat — a role conceptually distinct from perceived risk or autonomy threat, which describe properties of the technology itself rather than the individual's evaluative response to it. Therefore:

*H2: Perceived ease of AI use moderates the relationship between vicarious exposure and AI job replacement anxiety, such that the higher the perceived ease of AI use, the stronger the relationship.*

*H3: Perceived usefulness of AI moderates the relationship between vicarious exposure and AI job replacement anxiety, such that the higher the perceived usefulness of AI, the stronger the relationship.*

A growing body of evidence suggests that stable individual differences shape baseline vulnerability to technology-related anxiety, yet situational perceptions of control remain critical triggers of such responses. Core self-evaluations (CSE) capture a broad, trait-like sense of self-worth, efficacy, and emotional stability [49] and have been shown to buffer negative reactions to technological change, including affective job insecurity related to automation [50]. Relatedly, neuroticism is positively associated with AI anxiety [20], but this relationship appears to be attenuated when broader self-evaluative resources such as CSE are taken into account. Similarly, self-efficacy—a central component of CSE—has been shown to reduce job stress linked to AI adoption [51] and to buffer the adverse effects of job insecurity on psychological safety [52]. At the same time, perceptions of control represent a more proximal, situational factor that can elicit anxiety even among individuals with relatively strong dispositional resources. Empirical work demonstrates that a perceived lack of control over life events is associated with heightened fear of automation [15], underscoring the role of control-related appraisals in shaping technology-related fears. Building on this distinction between trait-like buffers and state-like triggers, we propose that perceived AI-over-human control will be associated with AI anxiety above and beyond core self-evaluations, reflecting the capacity of situational control perceptions to activate anxiety responses even when baseline predispositions are taken into account.

*H4: Perceived AI control will be associated with AI anxiety after controlling for core self-evaluations*.

## Study 1

### Method

#### *Sample and sampling*

Eligibility criteria were defined a priori. Participants had to (a) be adults and (b) be active on the labor market or have recent work experience, ensuring current exposure to work-related contexts relevant to the study variables. Exclusion criteria were limited to procedural and data-quality issues. Participants were removed if they did not complete the study or if technical errors compromised data integrity. Eight individuals were excluded on this basis, and one additional outlier was removed prior to regression analyses according to pre-specified diagnostic criteria.

The intended sample size (N = 325) was determined a priori using G*Power 3.1.9.7, assuming $\alpha = .01$, $\beta = .01$, and a small population effect size. A total of 325 individuals participated in the experimental procedure. After exclusions, the analytic sample comprised 316 participants. Demographic and control characteristics are presented in Table 1, disaggregated by randomized condition to permit assessment of baseline equivalence.

The study employed a randomized recruitment procedure within an accessible correspondence database. Individuals were invited in randomized blocks of 100 potential participants at a time until all study slots were filled. Recruitment was supplemented by distributing information about the study opportunity via social media groups associated with local communities and professional working groups. Interested individuals first completed an application form, after which eligible respondents were invited to participate until the intended sample size was achieved.

An effect-size sensitivity analysis was conducted in G*Power 3.1.9.7 to evaluate statistical sensitivity further. Assuming $\alpha = .01$, $\beta = .01$, a multiple regression model with 10 predictors, and N = 316 (after exclusion of outlier), the minimum detectable effect was $R^2 = .072$. These results indicate that the study was sufficiently powered to detect small-to-moderate effects under conservative error thresholds.

#### *Measures*

*Perceived AI-over-Human Control*. Prior to the experimental manipulation, the variable was measured using two items designed to capture baseline perceptions of AI autonomy ( “Artificial intelligence operates independently and makes decisions without human intervention” and “The functioning of artificial intelligence is entirely dependent on humans—both its rules and the decisions it makes”) to capture the baseline level of the independent variable. Statements were rated on a 5-point Likert scale. As expected for a very short scale, internal consistency was modest, Cronbach’s $\alpha = .62$; Feldt’s 95% CI [.55, .68]. Following the experimental manipulation, perceived AI-over-human control was reassessed using four items to evaluate the effectiveness of the manipulation, with two items reflecting AI control over humans and two reflecting human control over AI (e.g., “Decisions made by artificial intelligence are controlled by artificial intelligence”), drawn from Zhan et al. [21] and Molina and Sundar [53], and rated on the same 5-point Likert scale. This post-test measure demonstrated acceptable internal consistency, Cronbach's $\alpha$ = .63; Feldt's 95% CI [.56, .69].

*Perceived Ease of AI Use.* This variable was assessed at the end of the study to check the moderator level manipulation, using four items drawn from prior studies on the Technology Acceptance Model [54,55]. The items were reworded to refer specifically to artificial intelligence technologies while preserving their original meaning, thereby capturing the perceived effortlessness of interacting with AI (e.g., "Interaction with AI is intuitive and easy to understand"). Responses were recorded on a 5-point Likert scale, and the measure demonstrated good internal consistency, Cronbach's $\alpha$ = .82; Feldt's 95% CI [.79, .85].

*Perceived Usefulness of AI*. The variable was assessed at the end of the study to verify the moderator level manipulation, using four items drawn from prior studies on the Technology Acceptance Model [54,55]. *The items were reworded to refer specifically to artificial intelligence technologies while preserving their original meaning, thereby capturing beliefs about the extent to which AI enhances performance across work-related contexts (e.g., "Using AI would enhance one's performance across various types of work"). Responses were recorded on a 5-point Likert scale, and the measure demonstrated excellent internal consistency, Cronbach's* $\alpha$ = .89; Feldt's 95% CI [.87, .91].

*AI job replacement anxiety.* The dependent variable was measured using a six-item subscale developed by Wang and Wang [13]. Participants responded on a 5-point Likert scale. A sample item is "I am afraid that an AI technique or product may replace humans." The scale demonstrated good internal consistency in the present study, Cronbach's $\alpha$ = .83; Feldt's 95% CI [.80, .86].

*CSE* were assessed using the 12-item Core Self-Evaluations Scale [CSES; 49]. Responses were recorded on a 5-point Likert scale. A sample item is "I am confident I get the success I deserve in life." The scale demonstrated acceptable internal consistency in the present study, Cronbach's $\alpha$ = .77; Feldt's 95% CI [.73, .81].

*General negative attitudes toward artificial intelligence* were included as a control variable to account for baseline evaluative biases that may amplify anxiety responses to AI-related content. This construct was measured using four items selected from the General Attitudes toward Artificial Intelligence Scale [GAAIS; 56], excluding items directly related to job replacement or employment outcomes. Responses were recorded on a 5-point Likert scale. A sample item is "I think artificially intelligent systems make many errors." The scale demonstrated acceptable internal consistency, Cronbach's $\alpha$ = .74; Feldt's 95% CI [.69, .78].

*Trait anxiety* was included as a control variable to account for baseline anxiety levels that could influence responses to the experimental manipulation. It was measured using three items drawn from the cognitive component of the trait part of the State–Trait Inventory for Cognitive and Somatic Anxiety [STICSA; 57]. Participants indicated how frequently they experience each state on a 4-point scale. A sample item is "I picture some future misfortune." The scale demonstrated good internal consistency, Cronbach's $\alpha$ = .81; Feldt's 95% CI [.78, .84].

To account for potential confounds related to demographic background, employment context, and prior labor-market experiences, several control variables were measured. Demographic controls included *gender* and *age*. *Educational attainment* was measured categorically, distinguishing primary or lower, secondary, and higher education. Work-related characteristics were assessed by measuring *total job tenure* (in years), *type of work performed* (office-based or intellectual, physical, care-related, or hybrid), and *industry of employment*. Industries were recorded to capture sectors with varying exposure to AI-related automation, including customer service, office administration, media and marketing, programming and data analysis, finance and accounting, as well as an open "other" category. To capture recent labor-market vulnerability, participants reported whether they had been *fired from a job in the past 12 months*, whether they had experienced *difficulties finding a job* during that period, and whether they were *the sole provider* of income in their household. Finally, *prior exposure to artificial intelligence* in the workplace was assessed by asking whether participants currently or previously used AI at work (regularly, occasionally, only for private purposes, or not at all).

### *Data Collection Procedure*

The study employed a preregistered randomized between-subjects experimental method, conducted without modifications to the protocol approved by the institutional ethics committee and in accordance with the Declaration of Helsinki. Participants were randomly assigned to one of the experimental conditions in a 2 (Perceived control: AI vs. Human) × 2 (Perceived ease of use: Low vs. High) × 2 (Perceived usefulness: Low vs. High) design.

The study was conducted in a controlled laboratory setting using a dedicated testing platform displayed on computer workstations. The laboratory contained three separate workstations, each partitioned by black screens to ensure privacy. Each station was equipped with a laptop connected to an external monitor, keyboard, mouse, and wireless noise-cancelling headphones. Participants completed the experiment individually, with up to three participants tested simultaneously per session. Data collection took place between April 14 and May 23, 2025.

Each experimental session began with participants receiving detailed technical instructions, including information about the study duration and procedures. To preserve the integrity of the manipulation, participants were informed that the study examined "how artificial intelligence is communicated to the public." They were given time to review all study information displayed on the computer screen and to provide informed consent electronically. Following consent, the procedure was completed entirely independently, without further researcher involvement.

In subsequent stages, participants completed a survey assessing demographic characteristics, control variables, and the baseline level of the dependent variable. They then viewed three compilations of short user-generated social media videos, each lasting approximately six minutes. Each compilation consisted of 5–10 videos (20–60 seconds each) featuring individuals (e.g., experts or reporters) presenting opinions or information designed to induce perception that relates to the targeted level of the independent variable or moderators. The videos depicted diverse speakers in terms of gender and race and included subtitles in Polish and English. Materials were selected from existing social media platforms and included only videos that were directly relevant to the target construct (e.g., AI control over humans), neutral in emotional tone, and free from controversial content. Short-form social media videos were chosen because they represent a dominant mode of contemporary information consumption [58,59] and have been shown to elicit emotional responses in the context of AI-related content [60]. First, participants viewed compilations related to the moderators (perceived usefulness or perceived ease of use, in randomized order), followed by a compilation targeting perceived AI or human control. Detailed descriptions are provided below.

The compilations targeting *perceived AI control* versus *perceived human control* operationalized the independent variable by inducing contrasting perceptions of agency (AI over humans vs. humans over AI). Each compilation presented content designed to support one of the two experimental conditions. Video topics included corporate use of AI to influence or monitor individuals, AI supervision and workplace surveillance, the concept of technological singularity and its projected time horizon, risks associated with rapid and unsupervised AI scaling, decision autonomy, autonomous replication, and broader themes of societal control. In the human-control condition, analogous topics emphasized human oversight, regulation, and governance of AI systems. Importantly, none of the videos explicitly addressed job replacement.

The compilations targeting *perceived ease of use* manipulated the moderator at either a *high* or *low* level. In the high-ease condition, videos emphasized the simplicity of learning to use AI at work, low implementation costs, and straightforward development or application of AI tools to solve specific job-related problems. In the low-ease condition, analogous content highlighted complexity, higher implementation demands, and difficulties in effectively applying AI technologies in workplace contexts.

The compilations targeting *perceived usefulness* manipulated the second moderator at either a *high* or *low* level. In the high-usefulness condition, videos highlighted AI's capability to assist at work, solve specific job-related problems, and improve performance through concrete applications and technologies. In the low-usefulness condition, videos presented examples and descriptions of AI's limitations and shortcomings in real workplace environments.

Each participant viewed one compilation from each category, assigned using simple randomization. The randomization sequence was generated in a spreadsheet prior to data collection and used to allocate participants to conditions in the order of their enrollment. Participants were instructed to attend carefully to the videos, as they would subsequently evaluate them. After each video, they rated its comprehensibility, informational value, and relevance to the topic, reinforcing the cover story. Following exposure to all compilations, the dependent variable—AI replacement anxiety—was assessed. To further preserve the masking procedure, participants were informed that additional measures were being collected to understand their evaluations of the videos better. Accordingly, the AI replacement anxiety scale was preceded by a brief AI knowledge test. On the final survey page, post-manipulation measures of perceived AI control, perceived ease of use, and perceived usefulness were administered.

Upon completion of the study, debriefing information was displayed at each workstation, explaining the true purpose of the research and the rationale for the masking procedure. After all participants had finished, the researcher distributed the compensation and provided additional clarification upon request.

**Results**

Participants were distributed evenly across the eight experimental cells of the $2 \times 2 \times 2$ design, with group sizes ranging from n = 37 to n = 42. Mean AI job replacement anxiety was consistently higher in the AI-over-human control condition (Ms = 3.74–4.08) than in the human-control condition (Ms = 2.90–3.78), whereas differences associated with perceived AI ease of use were less pronounced and showed no clear systematic pattern at the descriptive level. In contrast, higher levels of perceived AI usefulness were generally associated with higher AI anxiety across conditions (Ms = 2.86-3.74 for low usefulness, and Ms = 3.75-4.26 for high). Descriptive statistics for AI anxiety and all control and manipulation check variables are presented in Table 2. AI job replacement anxiety was moderate to high (M = 3.64, SD = 0.89), with distributions showing no substantial deviations from normality. Perceptions of AI-over-human control differed between pre- and post-manipulation assessments (Ms = 2.81 and 2.24, respectively). Perceived usefulness of AI was high (M = 4.43), whereas perceived ease of use was moderately high (M = 3.83). Core self-evaluations, general negative attitudes toward AI, and trait anxiety showed expected variability. Correlations were generally small to moderate and in theoretically consistent directions, with no indications of problematic distributional properties.

Randomization checks were conducted using $\chi^2$ tests for categorical control variables and one-way ANOVAs for continuous control variables. No statistically significant differences between experimental conditions were observed with respect to gender, $\chi^2(14) = 15.14$, $p = .369$; having experienced difficulties finding a job in the past 12 months, $\chi^2(14) = 15.79$, $p = .327$; being the sole provider in the household, $\chi^2(7) = 6.86$, $p = .444$; having higher education, $\chi^2(7) = 6.57$, $p = .475$; type of work performed, $\chi^2(21) = 24.11$, $p = .288$; industry of employment, $\chi^2(35) = 46.22$, $p = .097$; prior experience with using AI at work, $\chi^2(21) = 31.08$, $p = .072$; or job tenure, $F(7, 309) = 2.00$, $p = .055$. Moreover, no differences between experimental conditions were observed for the psychological control variables, including general negative attitudes toward AI, $F(7, 309) = 1.36$, $p = .222$, and trait anxiety, $F(7, 309) = 0.74$, $p = .642$. However, a statistically significant difference was observed for having lost a job in the past 12 months, $\chi^2(7) = 16.03$, $p = .025$; however, despite a moderate association (Cramér's *V* = .225), this difference was considered substantively negligible due to the low overall number of such cases in the sample (N = 28). In addition, a statistically significant difference was observed for age, $F(7, 307) = 2.33$, $p = .025$; however, this difference was deemed substantively insignificant due to its small effect size, $\eta^2 = .05$.

Independent-samples *t* tests were conducted to examine the effectiveness of the experimental manipulations. As expected, no differences between conditions were observed for perceived AI control assessed prior to the manipulation, $t = 0.66$, $p = .509$, $d = 0.07$, indicating comparable baseline perceptions across groups. In contrast, perceived AI control assessed after the manipulation differed significantly between conditions, $t = -7.27$, $p < .001$, $d = -0.82$, suggesting that the manipulation successfully induced the intended perception of AI having control over people. Furthermore, participants exposed to different levels of the moderator manipulation differed significantly in their perceived ease of use of AI, $t = -10.10$, $p < .001$, $d = -1.14$, as well as in their perceived usefulness of AI, $t = -4.04$, $p = .013$, $d = -0.43$, measured at the end of the study, indicating that both moderator manipulations effectively influenced the targeted perceptions.

Multiple regression analysis was conducted to test the study hypotheses. The model was specified to include the main experimental predictors and their interactions derived from the theoretical framework. In addition, core self-evaluations, general negative attitudes toward AI, and trait anxiety were included as psychological control variables. Given minor but negligible differences observed in preliminary analyses, age and having lost a job in the past 12 months were also entered as covariates. The initial model was statistically significant, $F(10, 306)= 20.90$, $p < .001$, explaining 40.6% of the variance in AI anxiety (39% after adjusting for variables number). After estimating this initial regression model, influential observations were identified using standard diagnostic criteria with preregistered assumptions. Cases were flagged only when Cook's distance exceeded the conventional threshold (4/N) and was accompanied by either large standardized residuals (> 3) or high leverage values. Using this conservative, influence-based criterion, only one observation was identified and excluded from subsequent analyses. After excluding influential observations, the final regression model was re-estimated ($N = 316$) and showed good overall fit, $F(10, 305) = 22.00$, $p < .001$, explaining 41.9% of the variance in AI anxiety (40% after adjusting for variables number). These results are presented in Table 3. The observed proportion of explained variance exceeded the critical $R^2$ identified in the sensitivity analysis, indicating that the detected effects were sufficiently large to be reliably interpreted given the sample size and model complexity. Shapiro–Wilk tests indicated deviations from normality for both the outcome, $W = 0.97$, $p < .001$, and the residuals, $W = 0.99$, $p = .024$; however, visual inspection of Q–Q plots showed close adherence to the normal distribution with only minor deviations in the extreme tails. Inspection of residuals-versus-fitted plots revealed no systematic patterns or funnel-shaped dispersion, consistent with the non-significant Breusch–Pagan test, $\chi^2 = 13.30$, $p = .207$, and Goldfeld–Quandt test, $F = 0.76$, $p = .951$, indicating no evidence of heteroskedasticity. Linearity and independence of observations were supported by the Rainbow test, $F = 0.88$, $p = .793$, and the Durbin–Watson statistic, $DW = 2.07$, $p = .669$. To ensure robust inference given minor deviations from normality, all regression coefficients were therefore estimated using heteroskedasticity-robust (HC3) standard errors.

Exposure to content emphasizing perception of AI having control, compared with human control, was associated with higher AI job replacement anxiety, $b = 0.96$, $SE = 0.15$, $t = 6.26$, $p < .001$, providing support for H1. The moderating role of perceived AI ease of use proposed in H2 was not supported, as the interaction between exposure condition and ease of use was not significant, $b = -0.07$, $SE = 0.18$, $t = -0.37$, $p = .710$. The interaction between exposure condition and perceived AI usefulness was significant, $b = -0.57$, $SE = 0.18$, $t = -3.21$, $p = .001$, however, its direction was opposite to that predicted by H3, indicating that the relationship between exposure to AI control cues and AI anxiety weakened as perceived usefulness increased. Beyond this interaction, perceived AI usefulness was also independently and positively associated with AI anxiety, $b = 1.00$, $SE = 0.12$, $t = 8.08$, $p < .001$, whereas perceived AI ease of use showed no independent association, $b = 0.04$, $SE = 0.13$, $t = 0.28$, $p = .777$. Support for H4 was observed in that perceived control of AI remained a significant predictor of AI anxiety after controlling for stable psychological characteristics and attitudes. Among these factors, general negative attitudes toward AI were positively related to AI anxiety, $b = 0.32$, $SE = 0.05$, $t = 6.91$, $p < .001$, whereas core self-evaluations, $b = -0.08$, $SE = 0.06$, $t = -1.44$, $p = .152$, and trait anxiety, $b = -0.01$, $SE = 0.06$, $t = -0.14$, $p = .887$, were not significantly associated with the outcome. Importantly, age and having lost a job in the past 12 months, which had shown minor between-condition differences in preliminary analyses, were not significantly associated with AI anxiety (age: $b = -0.01$, $SE = 0.01$, $t = -1.44$, $p = .151$; job loss: $b = 0.07$, $SE = 0.16$, $t = 0.44$, $p = .661$), and their inclusion did not alter the pattern of results.

Follow-up simple-effects analyses were conducted to clarify the interaction between perception of AI having control and perceived AI usefulness. When perceived AI usefulness was low, AI control perception (AI- vs. human-agent control) was associated with substantially higher AI anxiety, $b = 0.92$, $SE = 0.13$, $t(306) = 7.31$, $p < .001$. When perceived AI usefulness was high, this difference remained statistically significant but was smaller in magnitude, $b = 0.33$, $SE = 0.13$, $t(306) = 2.59$, $p = .010$, indicating that the moderation effect was significant but operated in the opposite direction to that predicted by H3. Inspection of estimated marginal means (see Figure 1 and Table 2) revealed that AI anxiety reached its highest absolute level under conditions of AI control perception combined with high perceived usefulness ($M = 4.17$), while the lowest level was observed under conditions of human-over-AI control perception and low perceived usefulness ($M = 2.88$). This pattern reflects the strong independent association between perceived usefulness and anxiety: higher perceived usefulness elevated anxiety across both exposure conditions, thereby raising the baseline against which the AI–human agency contrast operated and attenuating its relative magnitude without reversing its direction. Taken together, these findings indicate that H3 was not supported, i.e., the pre-registered directional prediction was disconfirmed. The following pattern is nonetheless interpretable post-hoc: perceived usefulness consistently amplified absolute levels of AI job replacement anxiety while simultaneously compressing the relative effect of AI control narratives.

## Study 2

All variables specified in the proposed model were measured, with constructs that were experimentally manipulated in Study 1 operationalized as observed variables in the present study. The purpose of Study 2 was to examine the model's external validity through a large-scale observational design in which all focal variables were assessed rather than experimentally manipulated.

Because Study 2 operationalizes perceived AI-over-human control as a measured attitude rather than an experimentally manipulated perception, it cannot establish causality or directly replicate the mechanism tested in Study 1. Its purpose is to examine whether the associations identified experimentally generalize to a naturalistic setting with a broader sample, not to provide a second causal test of the model.

## Method

### *Sample and sampling*

Eligibility criteria were defined a priori. Participants were required to (a) be at least 18 years old, (b) be currently employed, and (c) not be platform-based workers (e.g., Uber drivers).

Participants were recruited through a research panel using random sampling procedures. Of the 1,715 individuals who initiated the survey, 1,423 completed it. After excluding respondents who did not pass screening questions, 1,218 cases remained. Following the removal of participants who failed attention checks, the final analytic sample comprised 1,019 individuals. Demographic characteristics are presented in Table 1.

An effect size sensitivity analysis was conducted in G*Power 3.1.9.7 for linear multiple regression to determine the minimum detectable effect given the available sample size. Assuming a one-tailed test, $\alpha = .01$, and power = .99, and a total sample size of N = 995 (after exclusion of statistical outliers), the analysis yielded a critical $R^2$ of .017. This result indicates that the study was sufficiently powered to detect very small effects in the tested regression models under conservative statistical assumptions.

### *Measures*

The same measures as in Study 1 were used to assess the dependent variable (AI replacement anxiety), CSE, and demographic and control variables. These included gender, age, educational attainment, total job tenure, type of work performed, industry of employment, job loss within the past 12 months, difficulties in finding employment, sole-provider status, and prior experience with AI at work. The following variables were assessed differently from Study 1, as they were operationalized as observed rather than experimentally manipulated.

*Perceived AI-over-human control* was measured using 10 items, with half reflecting AI control over humans and the other half reflecting human control over AI (e.g., “Decisions made by artificial intelligence are controlled by artificial intelligence”), drawn from Zhan et al. [21] and Molina and Sundar [53], and rated on the same 5-point Likert scale. This post-test measure demonstrated acceptable internal consistency, Cronbach’s $\alpha = .84$; Feldt’s 95% CI [.82, .85].

*Perceived Ease of AI Use* was assessed using six items drawn from prior studies on the Technology Acceptance Model [54,55]. The items were reworded to refer specifically to artificial intelligence technologies while preserving their original meaning, thereby capturing the perceived effortlessness of interacting with AI (e.g., “Interaction with AI is intuitive and easy to understand”). Responses were recorded on a 5-point Likert scale, and the measure demonstrated excellent internal consistency, Cronbach’s $\alpha = .91$; Feldt’s 95% CI [.90, .92].

*Perceived Usefulness of AI* was assessed using six items drawn from prior studies on the Technology Acceptance Model [54,55]. The items were reworded to refer specifically to artificial intelligence technologies while preserving their original meaning, thereby capturing beliefs about the extent to which AI enhances performance across work-related contexts (e.g., "Using AI would enhance one's performance across various types of work"). Responses were recorded on a 5-point Likert scale, and the measure demonstrated excellent internal consistency, Cronbach's $\alpha = .94$; Feldt's 95% CI [.94, .95].

### ***Data Collection Procedure***

The study was conducted without modifications to the protocol approved by the institutional ethics committee and in accordance with the Declaration of Helsinki. Participants recruited via the research panel received a link to an online questionnaire. After reading the brief study instructions, they provided informed consent electronically before proceeding. Screening questions were administered at the outset; individuals who did not meet the eligibility criteria were prevented from continuing.

Eligible participants first completed demographic and control measures, followed by scales assessing the moderators (perceived ease of use and perceived usefulness of AI), perceived AI-over-human control, and AI replacement anxiety. Data collection was conducted between June 3 and July 10, 2025. The average survey completion time was 13 minutes and 57 seconds.

To ensure data quality, three attention checks were embedded throughout the survey. Failure to respond correctly resulted in exclusion from the analyses. An example item was: "In this row, select *disagree*." Participants were compensated non-monetarily through the research panel to reduce the risk of automated or non-human responses that monetary rewards may incentivize.

## Results

Table 4 presents descriptive statistics and zero-order correlations for the main study variables in Study 2. Means, standard deviations, 95% confidence intervals, skewness, kurtosis, and intercorrelations are reported. AI anxiety was positively correlated with perceived AI-over-human control, perceived ease of use, and perceived usefulness of AI, while CSE showed small but significant associations with the focal constructs.

Because all study variables were measured at the same time using self-reported questionnaires, the potential influence of common method bias was formally assessed. Harman's single-factor test was conducted using exploratory factor analysis on all item-level measures. The analysis employed minimum residual extraction, with the number of factors fixed to one. Sampling adequacy was high, MSA = .92, and Bartlett's test of sphericity was significant, $\chi^2(780) = 25825.81$, $p < .001$, indicating that the correlation matrix was suitable for factor analysis. The single-factor solution accounted for 27% of the total variance, remaining well below the 50% threshold suggested by Podsakoff et al. [61]. Together, these results suggest that common method variance did not dominate the covariance structure and is therefore unlikely to influence the interpretation of the study's findings substantially.

To test the study hypotheses, a linear regression model was estimated with AI job replacement anxiety as the dependent variable, perceived AI-over-human control as the focal predictor, its interactions with perceived ease of AI use and perceived usefulness of AI, and CSE as a control variable. After the initial model estimation, influential observations were identified using a preregistered, conservative, multi-criterion diagnostic procedure. As in the Study 1 case, observations were flagged only when Cook's distance exceeded the conventional threshold (4/N) and was accompanied by either large standardized residuals (> 3) or high leverage values. Applying these criteria, 24 observations (less than 2.5% of the sample) were identified as potentially influential. The regression model was then re-estimated after excluding these cases. All effects retained the same direction and statistical significance as in the full-sample model. Accordingly, results are reported for the cleaned dataset, as it yields more stable parameter estimates while leading to identical substantive conclusions.

Following data cleaning and re-estimation of the regression model, standard model assumptions and diagnostic checks were evaluated to ensure the robustness of subsequent inferences. The overall model was statistically significant, $F(4, 990) = 64.53$, $p < .001$, explaining a substantial proportion of variance in AI job replacement anxiety, $R^2 = .282$, adjusted $R^2 = .277$. Notably, the observed $R^2$ exceeded the critical $R^2$ derived from the sensitivity analysis, supporting the robustness and interpretability of the regression results. Results of the regression analysis are presented in Table 5. Rainbow tests of linearity, $F = .919$, $p = .827$, and Durbin-Watson test for independence of observations, $DW = 2.03$, $p = .70$, indicated no violations of these assumptions. Although Shapiro–Wilk tests suggested deviations from normality for both the outcome, $W = .99$, $p < .001$, and residuals, $W = .99$, $p < .001$, this was expected given the large sample size (N = 995) and did not materially affect estimation. Importantly, no evidence of heteroskedasticity was detected, $GQ = 1.01$, $p = .463$. Multicollinearity was not a concern, as variance inflation factors were low for all predictors, $VIFs \leq 1.30$. Taken together, these diagnostics indicate that the model provided a reliable basis for coefficient estimation and hypothesis testing. As for Study 1, all regression coefficients, standard errors, confidence intervals, and significance tests were estimated using HC3 heteroskedasticity-consistent robust standard errors.

Results from the regression analysis (Table 5) supported the central predictions of the study. Consistent with H1, perceived AI-over-human control was positively associated with AI job replacement anxiety, b = 0.55, SE = 0.03, 95% CI [0.49, 0.62], p < .001, indicating that vicarious exposure to narratives emphasizing AI agentic control increased replacement-related anxiety. Supporting H4, this effect remained robust after controlling for CSE, which were themselves negatively related to AI job replacement anxiety, b = −0.12, SE = 0.03, 95% CI [−0.17, −0.07], p < .001. However, the moderation hypotheses once again yielded mixed support. Contrary to H2, the interaction between perceived AI-over-human control and perceived ease of use was not statistically significant, b = 0.05, SE = 0.03, 95% CI [−0.02, 0.11], p = .143, indicating that perceived ease of use did not condition the effect of vicarious exposure on AI job replacement anxiety. In contrast, perceived usefulness significantly moderated the relationship between AI-over-human control and job replacement anxiety, b = −0.07, SE = 0.03, 95% CI [−0.13, −0.01], p = .03. This interaction indicates that the strength of the association between vicarious exposure and AI job replacement anxiety varied as a function of perceived usefulness of AI. In addition, perceived ease of use was not directly associated with AI job replacement anxiety, $b = 0.04$, $SE = 0.03$, 95% CI [−0.03, 0.10], $p = .27$, whereas perceived usefulness showed a significant positive association, $b = 0.10$, $SE = 0.04$, 95% CI [0.03, 0.17], $p = .007$.

To probe the significant interaction between perceived AI-over-human control and perceived usefulness, simple slopes analyses were conducted using HC3 robust standard errors (Figure 2). When perceived usefulness of AI was low (−1 SD), perceived AI-over-human control was strongly and positively associated with AI job replacement anxiety, $b = 0.62$, $SE = 0.04$, $t = 14.40$, $p < .001$. This association remained significant at the mean level of perceived usefulness, $b = 0.55$, $SE = 0.03$, $t = 17.00$, $p < .001$, and remained significant but weaker when perceived usefulness was high (+1 SD), $b = 0.48$, $SE = 0.05$, $t = 10.10$, $p < .001$. These results replicate the pattern observed in Study 1: the moderation effect was significant but operated in the opposite direction to that predicted by H3, with the association between perceived AI-over-human control and job replacement anxiety attenuating as perceived usefulness increased. Critically, this attenuation reflects the strong independent association between perceived usefulness and anxiety rather than a suppression of the control-exposure effect. Higher perceived usefulness elevated anxiety across all levels of perceived AI-over-human control, raising the baseline and thereby compressing the relative magnitude of the control narrative effect without reversing its direction. Taken together, these findings provide convergent evidence across both studies that although H3 was not supported in its predicted direction, perceived usefulness consistently amplified absolute levels of AI job replacement anxiety while attenuating the relative increase attributable to AI control narratives.

## Discussion

The present research aimed to clarify how AI job replacement anxiety emerges by integrating insights from Integrated Fear Acquisition Theory and the Technology Acceptance Model and by moving beyond predominantly correlational accounts. Across two complementary studies, an experimental investigation (Study 1; $N = 316$) and a large-scale observational study (Study 2; $N = 995$), we examined whether vicarious exposure to narratives emphasizing AI-over-human control increases job replacement anxiety and under which conditions this effect is amplified. Consistent with Hypothesis 1, both studies demonstrated that stronger perceptions of AI-over-human control were associated with higher levels of AI job replacement anxiety. Hypothesis 2 was not supported, as perceived ease of AI use did not moderate this relationship in either study. In contrast, Hypothesis 3 was not supported in its predicted direction: although perceived usefulness of AI significantly moderated the association between perceived AI control and job replacement anxiety across both studies, the interaction operated in the opposite direction to that predicted, with the effect of AI control narratives on anxiety attenuating rather than strengthening as perceived usefulness increased. Nevertheless, perceived usefulness was consistently associated with elevated absolute levels of job replacement anxiety across both exposure conditions, such that anxiety reached its highest level when AI was perceived as both highly useful and dominant. Finally, Hypothesis 4 was supported in both studies, showing that perceived AI control remained a significant predictor of AI job replacement anxiety after controlling for core self-evaluations.

Together, these findings provide converging experimental and observational evidence for a causal, theory-informed model of AI job replacement anxiety emergence.

**Theoretical contributions**

The present study advances theory by identifying perceived AI-over-human control as a proximal psychological mechanism underlying AI job replacement anxiety. By integrating Integrated Fear Acquisition Theory with constructs from the Technology Acceptance Model [12,54], it shows that anxiety is driven by appraisals of agency and control shifts, rather than by technological complexity or adoption likelihood alone. Importantly, the findings extend existing accounts of vicarious exposure [12,21] by demonstrating that generalized narratives portraying AI as an autonomous agent are sufficient to activate fear responses, even in the absence of explicit job loss events, thereby broadening theoretical models of how AI-related anxieties emerge before objective labor-market threats materialize.

Support for Hypothesis 1 highlights perceived AI agency as a relevant psychological component in contemporary accounts of AI job replacement anxiety. Across both studies, perceptions of AI operating autonomously functioned as a potent psychological threat cue, eliciting anxiety even in the absence of direct personal risk of displacement. This finding extends prior work by Li [12] by demonstrating that vicarious exposure to generalized narratives portraying AI as an autonomous agent is sufficient to elicit job replacement anxiety, even without direct exposure to job displacement. These results indicate that symbolic cues of agency and autonomy—rather than, e.g., mere technological complexity or novelty—constitute effective fear stimuli. Such cues are threatening because they imply irreversible shifts in control, asymmetrical power relations, and a diminishing relevance of human labor and judgment [16]. Importantly, this psychological mechanism operates in a context where the actual extent and timing of AI-driven job replacement remain uncertain, with empirical studies and expert reports offering mixed and often contradictory projections [e.g., 62–64]. Nevertheless, widespread conflation of computational autonomy with human-like autonomy fosters fears of AI developing free will and unpredictability, amplifying perceived loss of control. As narratives emphasizing AI agency and dominance are increasingly prevalent in both social and traditional media [7–9], the present findings suggest that such content constitutes a meaningful risk factor for workers' well-being by repeatedly activating a significant anxiety trigger grounded in perceived shifts of agency rather than in observable labor-market realities.

The present findings indicate that AI job replacement anxiety can be triggered by proximal psychological appraisals, not solely by broader structural labor-market conditions. More specifically, the results suggest that perceptions, particularly perceptions of AI agency and control, may contribute to anxiety responses beyond objective exposure to AI technologies or direct displacement experiences. This interpretation aligns with existing research showing that exposure to automation-related risks can trigger fear of automation, though not uniformly across workers [15]. The present study explains AI anxiety by identifying an additional, perceptual pathway through which AI job replacement anxiety may emerge. Prior research has largely emphasized structural and technological perspectives, linking fear of automation to job vulnerability, task routineness, and exposure to substitutive technologies [15,e.g., 52], as well as to perceived technological affordances [21,e.g., 65]. By contrast, the results point to perceptual and vicarious processes as an additional pathway, whereby exposure to narratives emphasizing AI agency may activate threat responses prior to observable labor-market consequences. In this sense, AI job replacement anxiety may develop not only as a reaction to observable technological change but also as an anticipatory response shaped by how AI is perceived, interpreted, and socially transmitted.

An important theoretical question concerns whether AI job replacement anxiety should be conceptualized primarily as a stable disposition or as a context-sensitive affective response. Yang and Sundar [66] argue for a trait-based view, drawing parallels with computer anxiety and showing that dispositional AI anxiety predicts momentary anxiety responses. In contrast, others conceptualize AI anxiety as a state-like affective reaction that inhibits interaction with AI, consistent with broader treatments of technological anxiety as situationally elicited [13,16]. The present findings align more closely with the latter perspective, insofar as anxiety responses were elicited under controlled experimental conditions and varied as a function of perceived AI-over-human control, with converging patterns observed in an observational sample. This suggests that perceptions of AI agency may operate as a state-like appraisal that activates anxiety in specific contexts, rather than reflecting a fixed individual disposition. At the same time, such state responses are embedded in broader structural conditions, indicating that AI job replacement anxiety reflects an interaction between situational appraisals and stable predispositions.

The moderation findings point to a clear asymmetry between perceived usefulness and perceived ease of use. While ease of use is a central construct in technology acceptance [22,54], it did not condition anxiety responses, suggesting that effort-related appraisals may operate differently when individuals are not merely users of a technology but potential subjects of its consequences, as is the case with AI [67]. Next, the pre-registered prediction that perceived usefulness would amplify the effect of AI control narratives was disconfirmed: although a significant interaction emerged across both studies, it operated in the opposite direction, with the effect of AI control narratives on anxiety attenuating rather than strengthening as perceived usefulness increased. Critically, this attenuation does not indicate that perceived usefulness buffered anxiety — a post-hoc interpretation, offered here as exploratory, points to a different mechanism. The attenuation reflects the strong and consistent independent association between perceived usefulness and job replacement anxiety: because perceived usefulness robustly elevated anxiety across both exposure conditions and regardless of AI control framing, it raised the baseline against which the narrative effect operated, thereby compressing its relative magnitude without diminishing its absolute impact. This pattern indicates that appraisals of value and functional impact, rather than assessments of feasibility or usability, might be more central for triggering threat responses in this context, amplifying absolute levels of job replacement anxiety above and beyond how AI agency is framed and communicated.

One plausible explanation for the anxiety-amplifying role of perceived usefulness is that individuals tend to ignore the incremental and computational nature of technological development, instead focusing on imagined end states in which AI appears powerful, autonomous, and socially consequential. This pattern is consistent with Johnson and Verdicchio's [16] argument that AI anxiety often stems from misinterpreting computational systems as autonomous agents, such that perceived usefulness increases anxiety by making AI-driven replacement seem justified and consequential, rather than by highlighting how easily it can be implemented. A complementary explanation can be drawn from risk perception research, which distinguishes between perceived severity of consequences and perceived probability of occurrence as separate drivers of affective responses [68,69]. Evidence across domains shows that emotional reactions to risk are often shaped more strongly by perceived severity than by likelihood, with individuals frequently displaying limited sensitivity to changes in hazard probability, especially under conditions of high perceived threat [70,71]. In this context, the perceived usefulness of AI may amplify anxiety by signaling how consequential replacement could be rather than how likely it is to occur. Consistent with this logic, research suggests that severity dominates risk appraisal for high-impact threats, whereas probability plays a larger role for minor risks [72]. Given that AI is often framed as an existential risk [13,27], perceived usefulness may therefore serve as a key driver of anxiety by heightening the perceived stakes of AI-driven labor displacement.

Support for H4 indicates that perceptions of AI-over-human control predict job replacement anxiety over and above CSE, underscoring the role of situational threat appraisals. While prior research links AI anxiety to dispositional factors such as neuroticism and self-efficacy [18,73,74], these traits appear insufficient to account for anxiety responses triggered by perceived shifts in agency. The persistence of the effect after controlling for CSE suggests that AI-over-human control is appraised as a qualitatively distinct threat, one that challenges individuals' perceived capacity to cope. In this sense, AI job replacement anxiety may arise less from who people are and more from what the technology is perceived to do, particularly when it is framed as operating beyond human control.

**Study limitations**

Several limitations should be noted. First, the temporal scope of the present studies was restricted to immediate anxiety responses, which precludes conclusions about the durability of AI job replacement anxiety or its longer-term psychological consequences. Future research should therefore employ longitudinal designs to examine the stability and temporal dynamics of AI job replacement anxiety over time.

Second, vicarious exposure was operationalized through curated media narratives, primarily in video-based formats; other modes of exposure, such as interpersonal communication or organizational messaging, may engage different psychological processes.

Third, although the findings are consistent with a state-like activation of anxiety, the designs cannot fully disentangle transient situational responses from more stable dispositional tendencies, leaving the trait–state distinction of AI job replacement anxiety an open question for future research.

Next, the modest internal consistency of the pre-manipulation measure of perceived AI-over-human control in Study 1 (Cronbach's $\alpha = .62$) may have introduced additional measurement error. This measure was used solely as a brief baseline assessment prior to the experimental manipulation; nevertheless, future research should aim for higher reliability even in short pre-experimental measures, particularly when assessing constructs central to perceived agency and control.

Finally, although Study 2 was designed to extend the external validity of the experimental findings, direct comparison across studies is limited by differences in construct operationalization. Perceived AI-over-human control was experimentally induced in Study 1 and assessed with two post-manipulation items, whereas in Study 2 it was measured as a naturally occurring attitude using a ten-item scale. Accordingly, the observational associations observed in Study 2 cannot be interpreted as causal, nor can they confirm that the same psychological mechanism is at work. The convergence between studies should therefore be understood as consistent with the proposed model rather than as independent replication of its causal structure.

## Conclusion

This research advances the theoretical understanding of AI job replacement anxiety by demonstrating that such anxiety can emerge through perceptual and vicarious mechanisms rather than direct technological exposure or objective employment threat. Across two studies, perceptions of AI-over-human control functioned as a proximal trigger of anxiety, with narratives emphasizing AI agency reliably eliciting job replacement concerns even in the absence of explicit displacement cues. By integrating Integrated Fear Acquisition Theory with technology acceptance perspectives, the findings clarify that perceived shifts in agency drive anxiety responses and that perceived usefulness amplifies fear by increasing the perceived severity of potential consequences. Together, these results position AI job replacement anxiety as a psychologically grounded response to how AI is perceived and communicated, underscoring the central role of agency-focused narratives in shaping emotional reactions to emerging technologies.

### Funding

This research was funded by the Initiative of Excellence – Research University programme (Adam Mickiewicz University, Poznan, Poland, decision no. IDUB 140/04/POB5/0019).

### Declarations

**Ethics approval and consent to participate**. This study was approved by the Ethics Committee for Research Projects of the Faculty of Psychology and Cognitive Science, Adam Mickiewicz University (decision No. 07/04/2025) and was conducted without modifications to the approved protocol and in accordance with the Declaration of Helsinki. Informed consent was obtained from all participants.

**Availability of data and materials**. Data and materials are openly available at the Open Science Framework: https://osf.io/yrjg4/. The study was pre-registered at: https://osf.io/83pmb/. No substantial discrepancies were identified between the registered protocol and the actual study conducted.

**Competing interests**. The authors declare no competing interests.

**Author contributions**. Conceptualization: J.G., T.C.-S.; Methodology: J.G., T.C.-S.; Software: J.G.; Formal analysis: J.G.; Investigation: J.G., M.K., K.S.; Resources: J.G.; Data curation: J.G.; Writing – original draft: J.G., M.K., K.S.; Writing – review & editing: J.G., M.K., K.S., T.C.-S.; Visualization: J.G.; Supervision: J.G., T.C.-S.; Project administration: J.G.; Funding acquisition: J.G.

**Acknowledgements**. The authors thank Professor Paweł Kleka of Adam Mickiewicz University for his valuable consultation and expertise in the development of the study design.

**Use of artificial intelligence tools**. A large language model (ChatGPT, OpenAI) was used exclusively for proofreading the final manuscript text. The authors reviewed and edited the content as needed and take full responsibility for the content of the published article.

## References

1. Frey, C. B. & Osborne, M. A. The future of employment: How susceptible are jobs to computerisation? *Technological Forecasting and Social Change* **114**, 254–280 (2017).
2. World Economic Forum. *Four Futures for Jobs in the New Economy: AI and Talent in 2030*. https://www.weforum.org/publications/four-futures-for-jobs-in-the-new-economy-ai-and-talent-in-2030/ (2026).
3. Goldman Sachs. *How Will AI Affect the Global Workforce?* https://www.goldmansachs.com/insights/articles/how-will-ai-affect-the-global-workforce (2025).
4. Massachusetts Institute of Technology. *Project Iceberg - Coordinating the Human-AI Future*. https://iceberg.mit.edu/ (2025).
5. Lange, J., Alper, A. & Lange, J. Americans fear AI permanently displacing workers, Reuters/Ipsos poll finds. *Reuters* (2025).
6. Grupa Pracuj.pl. *Pokolenie Z a AI na rynku pracy*. https://media.pracuj.pl/387392-paradoks-ai-38-mlodych-pracownikow-boi-sie-o-swoja-prace-mimo-ze-najlepiej-znaja-sztuczna-inteligencje (2025).
7. Qiu, S. S., Zhang, L., You, F. & Zhao, X. Unpacking media channel effects on AI perception: A network analysis of AI information exposure across channels, overload, literacy, and anxiety among Chinese users. *Computers in Human Behavior* **173**, 108790 (2025).
8. Appiah, E. & Htait, A. AI in the Public Eye: Analysing Social Media Sentiment and Opinion on Artificial Intelligence. in 1–15 (PMLR, 2025).
9. Deng, R. & Ahmed, S. Perceptions and paradigms: An analysis of AI framing in trending social media news. *Technology in Society* **81**, 102858 (2025).
10. Brougham, D. & Haar, J. Smart Technology, Artificial Intelligence, Robotics, and Algorithms (STARA): Employees' perceptions of our future workplace. *Journal of Management & Organization* **24**, 239–257 (2018).
11. Sha, C., Wang, L. & Pan, X. Does artificial intelligence (AI) anxiety increase employees' deviant behavior toward the organization? The role of emotional exhaustion and leadership support. *JPA* **35**, 207–214 (2025).
12. Li, J. & Huang, J.-S. Dimensions of artificial intelligence anxiety based on the integrated fear acquisition theory. *Technology in Society* **63**, 101410 (2020).
13. Wang, Y.-Y. & Wang, Y.-S. Development and validation of an artificial intelligence anxiety scale: an initial application in predicting motivated learning behavior. *Interactive Learning Environments* **30**, 619–634 (2022).
14. Kim, J. *et al.* AI Anxiety: a comprehensive analysis of psychological factors and interventions. *AI Ethics* **5**, 3993–4009 (2025).
15. Włoch, R., Śledziewska, K. & Rożynek, S. Who's afraid of automation? Examining determinants of fear of automation in six European countries. *Technology in Society* **81**, 102782 (2025).

16. Johnson, D. G. & Verdicchio, M. AI Anxiety. *Journal of the Association for Information Science and Technology* **68**, 2267–2270 (2017).
17. Ding, K. A Review Study on the Formation Factors of AI Anxiety and Its Impact on Employees. in 738–744 (Atlantis Press, 2025). doi:10.2991/978-94-6463-811-0_77.
18. Hajek, A. *et al.* Translation and validation of the artificial intelligence anxiety scale in German. *PLOS ONE* **20**, e0333073 (2025).
19. Alkhalifah, J. M., Bedaiwi, A. M., Shaikh, N., Seddiq, W. & Meo, S. A. Existential anxiety about artificial intelligence (AI)- is it the end of humanity era or a new chapter in the human revolution: questionnaire-based observational study. *Front. Psychiatry* **15**, (2024).
20. Stănescu, D. F. & Romașcanu, M. C. The influence of AI Anxiety and Neuroticism in Attitudes toward Artificial Intelligence. *European Journal of Sustainable Development* **13**, 191–191 (2024).
21. Zhan, E. S., Molina, M. D., Rheu, M. & Peng, W. What is There to Fear? Understanding Multi-Dimensional Fear of AI from a Technological Affordance Perspective. *International Journal of Human–Computer Interaction* **40**, 7127–7144 (2024).
22. Venkatesh, V. & Bala, H. Technology Acceptance Model 3 and a Research Agenda on Interventions. *Decision Sciences* **39**, 273–315 (2008).
23. Bangash, S. H., Khan, I., Husnain, G., Irfan, M. A. & Iqbal, A. Revolutionizing Healthcare with Smarter AI: In-depth Exploration of Advancements, Challenges, and Future Directions. *VFAST Transactions on Software Engineering* **12**, 152–168 (2024).
24. Benítez, M. & Parrado, E. Mirror, Mirror on the Wall: Which Jobs Will AI Replace After All?: A New Index of Occupational Exposure. *IDB Publications* https://doi.org/10.18235/0013125 (2024) doi:10.18235/0013125.
25. Maria, S., Purwinahyu, P., Fitriansyah, F., Rachmawaty, A. & Aini, R. N. Artificial Intelligence and Labor Markets: Analyzing Job Displacement and Creation. *International Journal of Engineering, Science and Information Technology* **5**, 290–296 (2025).
26. Wang, P. Three fundamental misconceptions of Artificial Intelligence. *Journal of Experimental & Theoretical Artificial Intelligence* **19**, 249–268 (2007).
27. Bostrom, N. *Superintelligence*. (Dunod, 2024).
28. Nyholm, S. A new control problem? Humanoid robots, artificial intelligence, and the value of control. *AI Ethics* **3**, 1229–1239 (2023).
29. Russell, S. Human-Compatible Artificial Intelligence. in *Human-Like Machine Intelligence* 3–23 (Oxford University Press, 2021). doi:10.1093/oso/9780198862536.003.0001.
30. Al-Smadi, S., Al-Smadi, F., Alzayyat, A. & Al-Shawabkeh, J. D. The Role of AI Anxiety and Attitudes Toward Artificial Intelligence in Shaping Healthcare Perceptions Among Jordanian Children. *The Open Nursing Journal* **19**, e18744346417980 (2025).
31. Britton, J. C., Lissek, S., Grillon, C., Norcross, M. A. & Pine, D. S. Development of anxiety: the role of threat appraisal and fear learning. *Depression and Anxiety* **28**, 5–17 (2011).
32. Qiao, Z., Pan, D., Hoid, D., van Winkel, R. & Li, X. When the approaching threat is uncertain: Dynamics of defensive motivation and attention in trait anxiety. *Psychophysiology* **59**, e14049 (2022).

33. Siegel, P., Cohen, B. & Warren, R. Nothing to Fear but Fear Itself: A Mechanistic Test of Unconscious Exposure. *Biological Psychiatry* **91**, 294–302 (2022).
34. Craske, M. G. Phobic fear and panic attacks: The same emotional states triggered by different cues? *Clinical Psychology Review* **11**, 599–620 (1991).
35. MacLeod, C. & Rutherford, E. M. Anxiety and the selective processing of emotional information: Mediating roles of awareness, trait and state variables, and personal relevance of stimu. *Behaviour Research and Therapy* **30**, 479–491 (1992).
36. Granić, A. & Marangunić, N. Technology acceptance model in educational context: A systematic literature review. *British Journal of Educational Technology* **50**, 2572–2593 (2019).
37. Kalayou, M. H., Endehabtu, B. F. & Tilahun, B. <p>The Applicability of the Modified Technology Acceptance Model (TAM) on the Sustainable Adoption of eHealth Systems in Resource-Limited Settings</p>. *JMDH* **13**, 1827–1837 (2020).
38. Tsai, T.-H., Lin, W.-Y., Chang, Y.-S., Chang, P.-C. & Lee, M.-Y. Technology anxiety and resistance to change behavioral study of a wearable cardiac warming system using an extended TAM for older adults. *PLOS ONE* **15**, e0227270 (2020).
39. Weng, F., Yang, R.-J., Ho, H.-J. & Su, H.-M. A TAM-Based Study of the Attitude towards Use Intention of Multimedia among School Teachers. *Applied System Innovation* **1**, 36 (2018).
40. Mishra, A., Shukla, A., Rana, N. P., Currie, W. L. & Dwivedi, Y. K. Re-examining post-acceptance model of information systems continuance: A revised theoretical model using MASEM approach. *International Journal of Information Management* **68**, 102571 (2023).
41. Nazari-Shirkouhi, S., Badizadeh, A., Dashtpeyma, M. & Ghodsi, R. A model to improve user acceptance of e-services in healthcare systems based on technology acceptance model: an empirical study. *J Ambient Intell Human Comput* **14**, 7919–7935 (2023).
42. Li, K. Determinants of College Students' Actual Use of AI-Based Systems: An Extension of the Technology Acceptance Model. *Sustainability* **15**, 5221 (2023).
43. Liu, M. *et al.* What influences consumer AI chatbot use intention? An application of the extended technology acceptance model. *Journal of Hospitality and Tourism Technology* **15**, 667–689 (2024).
44. Na, S., Heo, S., Han, S., Shin, Y. & Roh, Y. Acceptance Model of Artificial Intelligence (AI)-Based Technologies in Construction Firms: Applying the Technology Acceptance Model (TAM) in Combination with the Technology–Organisation–Environment (TOE) Framework. *Buildings* **12**, 90 (2022).
45. Saadé, R. G. & Kira, D. The Emotional State of Technology Acceptance. *Issues in Informing Science and Information Technology* **3**, (2006).
46. Faraoni, S. Persuasive Technology and computational manipulation: hypernudging out of mental self-determination. *Front. Artif. Intell.* **6**, (2023).
47. Federspiel, F., Mitchell, R., Asokan, A., Umana, C. & McCoy, D. Threats by artificial intelligence to human health and human existence. *BMJ Glob Health* **8**, (2023).
48. Prunkl, C. Human Autonomy at Risk? An Analysis of the Challenges from AI. *Minds & Machines* **34**, 26 (2024).

49. Judge, T. A., Erez, A., Bono, J. E. & Thoresen, C. J. The Core Self-Evaluations Scale: Development of a Measure. *Personnel Psychology* **56**, 303–331 (2003).
50. Walczok, M. & Bipp, T. Understanding the Emergence and Trajectory of Job Insecurity Due to Smart Technology, Artificial Intelligence, Robotics, and Automation. *Human Factors and Ergonomics in Manufacturing & Service Industries* **35**, e70000 (2025).
51. Kim, B. J. & Lee, J. The mental health implications of artificial intelligence adoption: the crucial role of self-efficacy. *Humanit Soc Sci Commun* **11**, 1561 (2024).
52. Kim, B. J. & Kim, M.-J. How artificial intelligence-induced job insecurity shapes knowledge dynamics: the mitigating role of artificial intelligence self-efficacy. *Journal of Innovation & Knowledge* **9**, 100590 (2024).
53. Molina, M. D. & Sundar, S. S. Does distrust in humans predict greater trust in AI? Role of individual differences in user responses to content moderation. *New Media & Society* **26**, 3638–3656 (2024).
54. Davis, F. D. Perceived Usefulness, Perceived Ease of Use, and User Acceptance of Information Technology. *MIS Quarterly* **13**, 319–340 (1989).
55. Venkatesh, V. & Davis, F. D. A Model of the Antecedents of Perceived Ease of Use: Development and Test. *Decision Sciences* **27**, 451–481 (1996).
56. Schepman, A. & Rodway, P. The General Attitudes towards Artificial Intelligence Scale (GAAIS): Confirmatory Validation and Associations with Personality, Corporate Distrust, and General Trust. *International Journal of Human–Computer Interaction* **39**, 2724–2741 (2023).
57. Ree, M. J., French, D., MacLeod, C. & Locke, V. Distinguishing Cognitive and Somatic Dimensions of State and Trait Anxiety: Development and Validation of the State-Trait Inventory for Cognitive and Somatic Anxiety (STICSA). *Behavioural and Cognitive Psychotherapy* **36**, 313–332 (2008).
58. Dixon, S. J. *U.S. Social Media Activities by Platform*. https://www.statista.com/statistics/200843/social-media-activities-by-platform-usa/ (2023).
59. Marcus, C. *Short Form Video Statistics and 2023 Marketing Trends*. https://www.colormatics.com/article/short-form-video-statistics-and-2020-marketing-trends/ (2023).
60. Klonek, F. & Parker, S. Does ai at work increase stress? Text mining social media about human–ai team processes and ai control. *Journal of Organizational Behavior* https://doi.org/10.1002/job.70000 (2025) doi:10.1002/job.70000.
61. Podsakoff, P. M., MacKenzie, S. B., Lee, J.-Y. & Podsakoff, N. P. Common method biases in behavioral research: A critical review of the literature and recommended remedies. *Journal of Applied Psychology* **88**, 879–903 (2003).
62. Malliaros, P. & Pacheco-Jaramillo, W. A. From Automation-Induced Job Loss to a Supervisory Economy? Preprint at https://doi.org/10.12688/f1000research.168512.1 (2025).
63. Occhipinti, J.-A. *et al.* Generative AI may create a socioeconomic tipping point through labour displacement. *Sci Rep* **15**, 26050 (2025).
64. Peiwen, C., Sulaiman, N. & Zhenglong, S. The Impact of Artificial Intelligence Application on Job Displacement and Creation: A Systematic Review. *IJRISS* **IX**, 2495–2517 (2025).

65. Upadhyay, N., Upadhyay, S. & Dwivedi, Y. K. Theorizing artificial intelligence acceptance and digital entrepreneurship model. *International Journal of Entrepreneurial Behavior & Research* **28**, 1138–1166 (2021).
66. Yang, H. & Sundar, S. S. AI anxiety: Explication and exploration of effect on state anxiety when interacting with AI doctors. *Computers in Human Behavior: Artificial Humans* **3**, 100128 (2025).
67. Williams, G. Y. & Lim, S. Psychology of AI: How AI impacts the way people feel, think, and behave. *Current Opinion in Psychology* **58**, 101835 (2024).
68. El-Toukhy, S. Parsing Susceptibility and Severity Dimensions of Health Risk Perceptions. *Journal of Health Communication* **20**, 499–511 (2015).
69. Wilson, R. S., Zwickle, A. & Walpole, H. Developing a Broadly Applicable Measure of Risk Perception. *Risk Analysis* **39**, 777–791 (2019).
70. Magnan, R. E., Gibson, L. P. & Bryan, A. D. Cognitive and Affective Risk Beliefs and their Association with Protective Health Behavior in Response to the Novel Health Threat of COVID-19. *J Behav Med* **44**, 285–295 (2021).
71. Weinstein, N. D. Perceived probability, perceived severity, and health-protective behavior. *Health Psychology* **19**, 65–74 (2000).
72. Siegrist, M. & Árvai, J. Risk Perception: Reflections on 40 Years of Research. *Risk Analysis* **40**, 2191–2206 (2020).
73. Kwak, Y., Ahn, J.-W. & Seo, Y. H. Influence of AI ethics awareness, attitude, anxiety, and self-efficacy on nursing students' behavioral intentions. *BMC Nurs* **21**, 267 (2022).
74. Sindermann, C. *et al.* Acceptance and Fear of Artificial Intelligence: associations with personality in a German and a Chinese sample. *Discov Psychol* **2**, 8 (2022).

## Tables and Figures

**Table 1**

Demographic characteristics of both studies participants

| Baseline characteristic | Study 1 | | | | | | | |
|---|---|---|---|---|---|---|---|---|
| | Human-agencies | | AI-agencies | | Overall | | Study 2 | |
| | *n* | % | *n* | % | *n* | % | *n* | % |
| **Gender** | | | | | | | | |
| Female | 109 | 34.7 | 107 | 34.1 | 216 | 68.8 | 652 | 55.8 |
| Male | 49 | 15.6 | 49 | 15.6 | 98 | 31.2 | 516 | 44.1 |
| **Highest educational level** | | | | | | | | |
| Primary education or lower | 0 | 0 | 2 | 0.6 | 2 | 0.6 | 20 | 1.41 |
| Secondary education | 80 | 25.5 | 88 | 28.0 | 168 | 53.5 | 434 | 34.01 |
| Higher education | 77 | 24.5 | 66 | 21.0 | 143 | 45.5 | 620 | 43.57 |
| PhD, MBA, or higher | 1 | 0.3 | 0 | 0.0 | 1 | 0.3 | 94 | 6.61 |
| **Employment type** | | | | | | | | |
| Office or intellectual work | 98 | 31.2 | 88 | 28.0 | 186 | 59.2 | 741 | 52.07 |
| Physical work | 26 | 8.3 | 28 | 8.9 | 54 | 17.2 | 312 | 21.93 |
| Care work | 20 | 6.4 | 19 | 6.1 | 39 | 12.4 | 57 | 4.01 |
| Combination | 14 | 4.5 | 21 | 6.7 | 35 | 11.1 | 108 | 7.59 |
| **Industry** | | | | | | | | |
| Customer service | 24 | 7.6 | 30 | 9.6 | 54 | 17.2 | 143 | 10.05 |
| Office administration | 16 | 5.1 | 16 | 5.1 | 32 | 10.2 | 232 | 16.30 |
| Media and marketing | 11 | 3.5 | 16 | 5.1 | 27 | 8.6 | 33 | 2.32 |
| IT and data analysis | 15 | 4.8 | 5 | 1.6 | 20 | 6.4 | 36 | 2.53 |
| Finance and accounting | 4 | 1.3 | 9 | 2.9 | 13 | 4.1 | 101 | 7.10 |
| Other | 88 | 28 | 80 | 25.5 | 168 | 53.5 | 673 | 47.29 |
| **Fired in last 12 month** | 14 | 4.5 | 14 | 4.5 | 28 | 8.9 | 60 | 4.22 |
| **Difficulties in job searching in last 12 month** | 59 | 29.5 | 51 | 25.5 | 110 | 55.0 | 136 | 9.56 |
| **Sole provider status** | 31 | 9.9 | 26 | 8.3 | 57 | 18.2 | 343 | 24.10 |
| **Ai utility** | | | | | | | | |
| Yes, regularly | 44 | 14.0 | 24 | 7.6 | 68 | 21.7 | 157 | 11.03 |
| Yes, occasionally | 48 | 15.3 | 37 | 11.8 | 85 | 27.1 | 349 | 24.53 |
| No, but I use AI privately | 55 | 17.5 | 71 | 22.6 | 126 | 40.1 | 304 | 21.36 |
| No, I have never used AI at work | 11 | 3.5 | 24 | 7.6 | 35 | 11.1 | 408 | 28.67 |

**Table 2**

Descriptive statistics (Study 1)

| Variable | Mean | 95% CI | | SD | Skew | Kurt | 1. | 2. | 3. | 4. | 5. | 6. | 7. |
|---|---|---|---|---|---|---|---|---|---|---|---|---|---|
| | | LL | UL | | | | | | | | | | |
| 1. AI job replacement anxiety | 3.64 | 3.54 | 3.74 | 0.89 | -0.45 | -0.39 | | | | | | | |
| 2. Perception of AI-over-Human Control (pre-manipulation) | 2.81 | 2.75 | 2.87 | 0.57 | -0.11 | 1.97 | .14* | | | | | | |
| 3. Perception of AI-over-Human Control (post-manipulation) | 2.24 | 2.16 | 2.32 | 0.7 | 0.17 | -0.23 | .21*** | .16* | | | | | |
| 4. Perceived ease of use of AI | 3.83 | 3.74 | 3.92 | 0.83 | -0.65 | -0.03 | .1 | .09 | .07 | | | | |
| 5. Perceived usefulness of AI | 4.43 | 4.36 | 4.5 | 0.59 | -0.96 | 0.65 | -.13* | .03 | -.08 | .16** | | | |
| 6. CSE | 3.24 | 3.16 | 3.32 | 0.68 | -0.35 | -0.1 | -.22*** | -.03 | .00 | -.07 | .04 | | |
| 7. General negative AI attitude | 3.53 | 3.45 | 3.61 | 0.68 | -0.24 | -0.24 | .39*** | -.02 | .05 | -.13* | -.23*** | -.24*** | |
| 8. Trait anxiety | 2.36 | 2.28 | 2.44 | 0.73 | 0.14 | -0.52 | .15** | .05 | -.02 | .08 | -.02 | -.63*** | .25*** |

*$p < .05$. **$p < .01$. ***$p < .001$.

**Table 3**

Results of the Regression Analysis Predicting AI Job Replacement Anxiety (Study 1)

| Effect | Estimate | *SE* | *95% CI* | | *p* | *VIF* |
|---|---|---|---|---|---|---|
| | | | LL | UL | | |
| Intercept | -0.6 | 0.18 | -0.95 | -0.26 | .001 | |
| AI-over-human control[a] | 0.96 | 0.15 | 0.66 | 1.25 | <.001 | 1.015 |
| Perceived ease of use[b] | 0.04 | 0.13 | -0.21 | 0.28 | .777 | 1.138 |
| Perceived usefulness[b] | 1 | 0.12 | 0.76 | 1.24 | <.001 | 1.146 |
| CSE | -0.08 | 0.06 | -0.19 | 0.03 | .152 | 1.312 |
| AI-over-human control × Perceived ease of use | -0.07 | 0.18 | -0.41 | 0.28 | .71 | |
| AI-over-human control × Perceived usefulness | -0.57 | 0.18 | -0.91 | -0.22 | .001 | |
| General negative AI attitude | 0.32 | 0.05 | 0.23 | 0.41 | <.001 | 1.059 |
| Trait anxiety | -0.01 | 0.06 | -0.12 | 0.11 | .887 | 1.355 |
| Age | -0.01 | 0.01 | -0.02 | 0 | .151 | 1.091 |
| Job loss in past 12 month | 0.07 | 0.16 | -0.24 | 0.38 | .661 | 1.023 |

| Statistic | Value | p |
|---|---|---|
| F | 22.018 | <.001 |
| $R^2$ | .419 | |
| adj. $R^2$ | .4 | |
| $W_y$ | .969 | <.001 |
| $W_{RES}$ | .99 | .024 |
| GQ | 0.761 | .951 |
| $F_{RAIN}$ | 0.876 | .793 |
| DW | 2.074 | .669 |

*Note*. N = 316. CI = confidence interval; LL = lower limit; UL = upper limit; Wy = Shapiro–Wilk test for normality of the outcome variable; $W_{RES}$ = Shapiro–Wilk test for normality of residuals; GQ = Goldfeld–Quandt test for heteroskedasticity; $F_{RAIN}$ = Rainbow test for linearity; DW = Durbin–Watson statistic for independence of observations.

[a]0 = human control, 1 = AI control perception. [b]0 = low, 1 = high level.

**Table 4**

Descriptive statistics (Study 2)

| Variable | Mean | 95% CI | | SD | Skew | Kurt | 1. | 2. | 3. | 4. |
|---|---|---|---|---|---|---|---|---|---|---|
| | | LL | UL | | | | | | | |
| 1. AI job replacement anxiety | 4.27 | 4.22 | 4.33 | 0.99 | 0.06 | 3.73 | | | | |
| 2. Perception of AI-over-Human Control | 3.30 | 3.26 | 3.34 | 0.71 | -0.08 | 5.67 | .60*** | | | |
| 3. Perceived ease of use of AI | 4.52 | 4.46 | 4.58 | 1.05 | -0.22 | 3.98 | .30*** | .35*** | | |
| 4. Perceived usefulness of AI | 4.71 | 4.65 | 4.77 | 1.10 | -0.43 | 4.07 | .27*** | .47*** | .61*** | |
| 5. CSE | 3.4 | 3.37 | 3.43 | 0.48 | -0.12 | 4.07 | .12*** | -.07* | .23*** | .22*** |

*$p < .05$. **$p < .01$. ***$p < .001$

**Table 5**

Results of the Regression Analysis Predicting AI Job Replacement Anxiety (Study 2)

| Effect | Estimate | *SE* | *95% CI* | | *p* | *VIF* |
|---|---|---|---|---|---|---|
| | | | LL | UL | | |
| Intercept | -0.02 | 0.03 | -0.07 | 0.03 | .457 | |
| AI-over-human control | 0.55 | 0.03 | 0.49 | 0.62 | <.001 | 1.003 |
| Perceived ease of use | 0.04 | 0.03 | -0.03 | 0.1 | .272 | 1.295 |
| Perceived usefulness | 0.1 | 0.04 | 0.03 | 0.17 | .007 | 1.27 |
| CSE | -0.12 | 0.03 | -0.17 | -0.07 | <.001 | 1.014 |
| AI-over-human control × Perceived ease of use | 0.05 | 0.03 | -0.02 | 0.11 | .143 | |
| AI-over-human control × Perceived usefulness | -0.07 | 0.03 | -0.13 | -0.01 | .03 | |
| | | | | | | |
| Statistic | Value | p | | | | |
| F | 64.531 | <.001 | | | | |
| $R^2$ | .282 | | | | | |
| adj. $R^2$ | .277 | | | | | |
| $W_y$ | .992 | <.001 | | | | |
| $W_{RES}$ | .991 | <.001 | | | | |
| GQ | 1.008 | .463 | | | | |
| $F_{RAIN}$ | 0.919 | .827 | | | | |
| DW | 2.033 | .7 | | | | |

*Note*. N = 995. CI = confidence interval; LL = lower limit; UL = upper limit; $W_y$ = Shapiro–Wilk test for normality of the outcome variable; $W_{RES}$ = Shapiro–Wilk test for normality of residuals; GQ = Goldfeld–Quandt test for heteroskedasticity; $F_{RAIN}$ = Rainbow test for linearity; DW = Durbin–Watson statistic for independence of observations.

**Figure 1**

AI replacement anxiety as a function of perceived AI control and perceived AI usefulness.

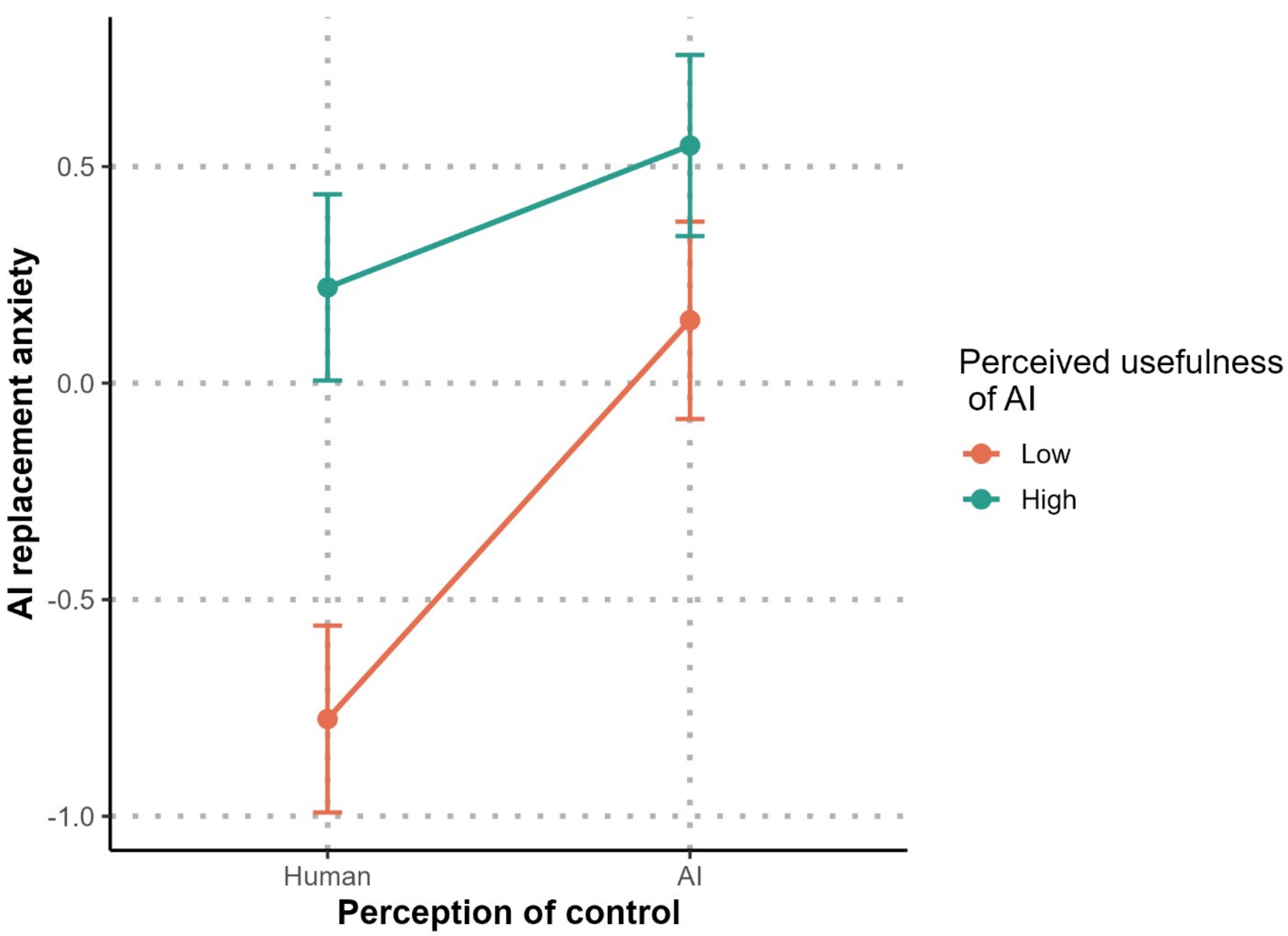

Note. Error bars represent 95% confidence intervals.

**Figure 2**

Conditional effects of perceived AI control on AI replacement anxiety as a function of perceived AI usefulness

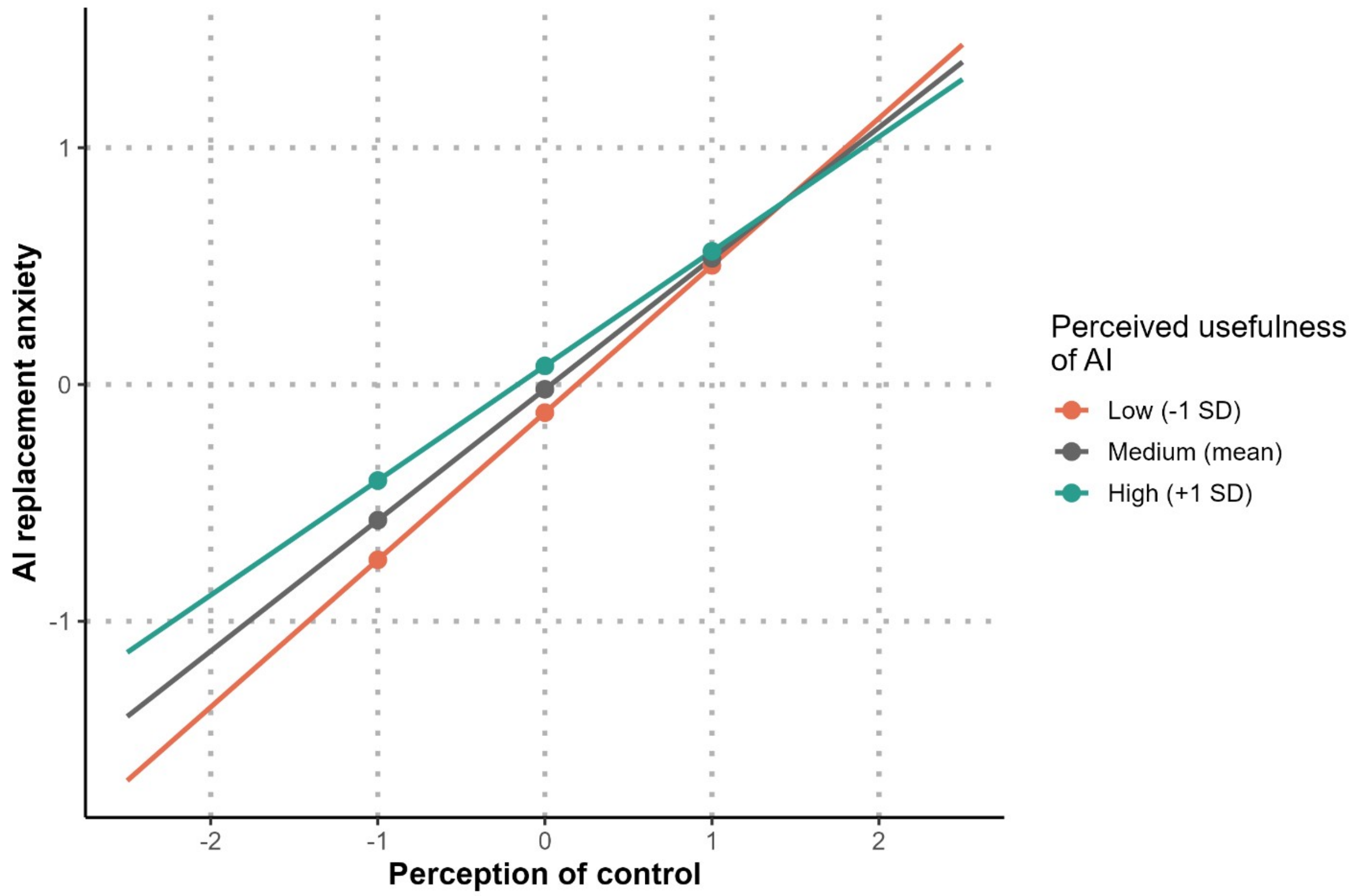